\documentclass[aps,pra,epsfig]{revtex4}

\usepackage[dvips]{graphics}
\usepackage{graphicx}
\usepackage{amsfonts}
\usepackage{amssymb}
\usepackage{amsmath}

\begin{document}

\title{Critical Temperature of 
Non-Interacting Bose Gases on Disordered Lattices}
\author{Luca Dell'Anna$^1$, Stefano Fantoni$^{1,2,3}$, Pasquale Sodano$^{4}$, and Andrea Trombettoni$^{1,2}$} 
\affiliation{$^1$ International School for Advanced Studies, 
Via Beirut 2/4, I-34014, Trieste, Italy\\
$^2$ INFN, Sezione di Trieste\\
$^3$ INFM, CNR-DEMOCRITOS National Supercomputing Center, Trieste, Italy\\
$^4$ Dipartimento di Fisica and INFN, Sezione di Perugia,  
Universit\`a di Perugia, Via A. Pascoli, 
I-06123, Perugia, Italy}

\begin{abstract}
For a non-interacting Bose gas on a lattice we compute the shift 
of the critical temperature for condensation when random-bond and onsite disorder are present. 
We evidence that the shift depends on the space dimensionality $D$ and the filling fraction $f$.
For $D \to \infty$ (infinite-range model), using results from
the theory of random matrices, we show that the shift of the critical
temperature is negative, depends on $f$, and vanishes only for large $f$. 
The connections with analogous results obtained for the spherical model are 
discussed. 
For $D=3$ we find that, for large $f$, the critical temperature $T_c$ is 
enhanced by disorder and that the relative shift does not sensibly 
depend on $f$; at variance, for small $f$, $T_c$ decreases 
in agreement with the results 
obtained for a Bose gas in the continuum. 
We also provide numerical estimates for the shift of the critical temperature due 
to disorder induced on a non-interacting Bose gas by a bichromatic incommensurate potential. 
\end{abstract}
\maketitle

\section{Introduction}

The ideal Bose gas (IBG) is a well studied topic in quantum statistical mechanics 
\cite{landau80,huang87,pathria96} since it provides invaluable qualitative insights on the properties of 
realistic bosonic systems. Its thermodynamics properties 
can be explicitly worked out and have been intensively studied both in the thermodynamical limit 
and in finite systems \cite{pathria98}. Nowadays, the study of the IBG has more than just academic 
interest since recent progresses in manipulating cold atomic systems 
\cite{pethick02,stringari03} allow for 
a reliable control of the relative strength of the 
kinetic and interaction energies: 
for instance, the possibilities of 
varying the scattering length through Feshbach resonances \cite{pethick02,stringari03} 
and 
superimposing optical lattices
(obtaining effective lattice Hamiltonians \cite{jaksch98,trombettoni01}) 
provide a tool for testing a Bose gas from the strongly interacting regime 
to the (almost) non-interacting one. 
In Ref.~\cite{roati07} it has been shown that it is possible to tune 
the $s$-wave scattering length $a$ of a $^{39}$Rb Bose gas 
with high precision, and set it almost to zero \cite{nota-dip}: 
this allows for long-lived Bloch oscillations in an optical lattice \cite{gustavsson08} 
and allows one to perform atom interferometry \cite{fattori08-b}. 
When an optical lattice is superimposed to an IBG, 
for large values of the laser power,
the system is described by a tight-binding bosonic model. 

The IBG has also been used to understand the effect of disorder 
\cite{lacour74} and the role played by confining, tailorable external potentials, relevant 
in investigations of trapped ultracold atoms \cite{pethick02,stringari03}. 
The effects of controllable or random impurities is again motivated by the huge progresses made  
on the manipulation of ultracold bosonic gases culminated in the recent experimental observation of
Anderson localization for matter waves in a random potential \cite{roati08,billy08}.
In a Bose gas, disorder can be induced both by a laser speckle
\cite{lye05,clement05} (i.e., a laser beam is shone through a diffusive plate and superimposed
to the trapping potential) or by an incommensurate bichromatic potential \cite{fallani07} (i.e.,
an auxiliary lattice is added to the main lattice, with an incommensurate ratio between the two frequencies).
Random on-site disorder in the strongly interacting limit induces a Bose glass phase
\cite{giamarchi88,fisher89}, whose characterization has been the object of many theoretical
\cite{gimperlein05,rey06,krutitsky06,roscilde07,sengupta07,buonsante07,deng08,roux08,morrison08,bissbort08}
and experimental \cite{fallani07} investigations.
Finally, ultracold atoms are a good experimental setup for studying superfluidity in random environments, 
which is an issue related to the adsorption of $^4$He in porous media 
\cite{kiewiet75,hertz79,fisher89,huang92,giorgini94,lopatin02,yukalov07}. 

The IBG is very interesting also in relation with statistical mechanics models: indeed, the IBG 
is in the universality class \cite{gunton68} of the spherical model (SM) \cite{berlin52}, 
which displays a phase transition in dimensions larger or equal than three. The SM 
is solvable in any dimension also with an external magnetic field 
\cite{berlin52,joyce72,mattis85}, 
and it provides an interesting laboratory to investigate a variety of topics, such as 
long-range interactions \cite{joyce66} and finite size scaling \cite{fisher72}. 
Furthermore, it is the $n \to \infty$ limit of the $O(n)$ model 
\cite{stanley68}, and, thus, is the leading term of $1/n$ expansions \cite{brezin93}. 
With disorder, the spin-glass SM with infinite-range interactions has been studied
intensively \cite{kosterlitz76,dedominicis06} and it has been used to analyze
the properties of relaxor ferroelectrics \cite{pirc99}.

When there is disorder, non-interacting bosons on a lattice are described by the Hamiltonian
\begin{equation}
\hat{H}=-t\sum_{i,j} \tilde{A}_{ij} \hat{b}^{\dag}_i \hat{b}_j +
\sum_i \epsilon_i \hat{b}^{\dag}_i \hat{b}_i,
\label{HAM}
\end{equation}
with
\begin{equation}
\tilde{A}_{ij}=A_{ij} (1+c_{ij}).
\label{A-tilde}
\end{equation}
In Eqs.~(\ref{HAM})-(\ref{A-tilde}), the lattice sites are denoted by $i,j$,
$t$ is the tunneling rate between neighboring sites, $A_{ij}$ is the adjacency matrix
characterizing the lattice (i.e., $A_{ij}=1$ if $i$ and $j$
are nearest neighbour sites, and $0$ if they are not) 
and $\hat{b}_i$, $\hat{b}_i^\dag$ are the bosonic operators. The number of sites
on a cubic lattice of dimension $D$ is $N_S=L^D$, while the total number of particles is $N_T$. 
The filling (i.e. the average number of particles per site) is
given by
\begin{equation}
f=\frac{N_T}{N_S}.
\label{filling}
\end{equation}
Random on-site disorder is accounted by $\epsilon_i$, while random-bond disorder is included 
in the definition of $A_{ij}$ through $c_{ij}$; in the following 
$\epsilon_i$ and $c_{ij}$ are regarded as random variables with vanishing average and 
variance $v_o^2t^2$ and $v^2$ respectively. We restrict ourself
to situations where $t_{ij}\equiv t \tilde A_{ij}>0$ ($v^2 \lesssim 1$), since they 
naturally arise when bosonic lattices are disordered
by inserting a speckle or by adding an incommensurate optical lattice: in these situations both 
random-bond disorder $c_{ij}$ and random on-site energies $\epsilon_i$ occur and, typically, $v_o \gtrsim v$. 
The inclusion of random-bond disorder allows also to make contact interesting properties of the spin-glass SM
\cite{kosterlitz76,dedominicis06}.

In this paper we investigate how the critical temperature of a lattice IBG is shifted by the
presence of disorder, elucidating the dependence of this shift on the filling. 
For the continuous Bose gas, the shift in the critical temperature has been derived
in Ref.~\cite{lopatin02}. In that paper it is shown that the critical temperature decreases with
disorder: as we shall see, the lattice computation reproduces this behaviour for small filling, but, 
on the lattice, different behaviours are possible as a result of the interplay between discreteness
and disorder. 
Indeed, on a lattice, there are 
two competing effects since disorder broadens the distribution of the
delocalized eigenstates of Hamiltonian (\ref{HAM}) and further localizes the ground-state 
wavefunction: the former (latter) effect tends to lower (increase) the critical temperature. 
For $D \to \infty$ (i.e., infinite-range model), the analysis can be carried analytically and 
one finds that the two effects exactly balance at large filling while, for small filling, the 
critical temperature decreases. For $D=3$ the
balance is obtained for intermediate filling while, at large (small) filling, the critical temperature
increases (decreases); the numerical coefficient that, for small disorder and large filling, 
determines the critical temperature increase is much smaller for random on-site disorder.

The plan of the paper is the following: in Section II we determine the critical temperature of 
non-interacting bosons on a three-dimensional lattice; there, we analyze the limit of large 
filling and the connection with the SM. Section III is devoted to the analysis of the 
infinite-range model without and with disorder. In Section IV we study 
three-dimensional lattices with random-bond disorder and random on-site energies; in addition, 
we exhibit numerical results 
for the shift of the critical temperature due to disorder induced by a bichromatic incommensurate 
potential. Our concluding remarks are in Section V.  

\section{Non-Interacting Bosons on a Lattice}

In this Section we determine the critical temperature for condensation of 
bosons hopping without disorder on a three dimensional lattice. 
We shall show also that, at large filling,  
the value of this temperature coincides with the one of a 
mean spherical model.

\subsection{Hopping Bosons without Disorder}

In the absence of disorder ($\epsilon_i=0$ and 
$c_{ij}=0$, $\forall i,j$), the Hamiltonian (\ref{HAM}) reduces to the boson-hopping Hamiltonian 
$\hat{H}^{(0)}$
\begin{equation}
\hat{H}^{(0)}=-t \sum_{i,j} A_{ij} \hat{b}^{\dag}_i \hat{b}_j.
\label{HAM_ord}
\end{equation}
By setting $\hat{d}_{{\bf k}}= N_S^{-1/2} 
\sum_j \hat{b}_j e^{i {\bf k} \cdot j}$, $\hat{H}^{(0)}$, $\hat{H}^{(0)}$ may be readily  
diagonalized; for $D=3$ one gets 
\begin{equation}
\hat{H}^{(0)}=\sum_{{\bf k}} E_{{\bf k}} \hat{d}^{\dag}_{{\bf k}} 
\hat{d}_{{\bf k}}
\label{HAM_ord_diag}
\end{equation}
with
\begin{equation}
E_{{\bf k}}=-2t \left( \cos{k_x} + \cos{k_y} + \cos{k_z} \right).
\label{autovalori}
\end{equation}
We assumed periodic boundary conditions, so that $k_\ell=2 \pi 
n_\ell /L $ with $n_\ell=0,\cdots,L-1$ ($\ell=x,y,z$).

In computing the grand-canonical partition function, one adds the chemical potential $\mu$ 
enforcing the conservation of the total number of particles; as a result, 
one replaces $\hat{H}^{(0)}$ with $\hat{K}^{(0)}=\hat{H}^{(0)}-\mu \hat{N}$, 
where $\hat{N}=\sum_i \hat{b}^{\dag}_i \hat{b}_i=\sum_{{\bf k}} 
\hat{d}^{\dag}_{{\bf k}} \hat{d}_{{\bf k}}$ is the operator counting the total number of particles. 
Requiring $\langle \hat{N} \rangle=N_T$ yields
\begin{equation}
\sum_{{\bf k}} \frac{1}{e^{\beta(E_{{\bf k}}-\mu)}-1}= N_T,
\label{condizione-somma}
\end{equation}
which fixes the chemical potential $\mu$ as a function of $T$ and $N_T$. In the thermodynamical limit, 
$N_S,N_T \to \infty$ at fixed filling $f$, 
Eq.~(\ref{condizione-somma}) reads
$\int_{BZ} \frac{d{\bf k}}{(2 \pi)^3}\frac{1}{e^{\beta(E_{{\bf k}}-\mu)}-1}= f$, 
where $\int_{BZ} d{\bf k} \cdots$ denotes 
the integral over the first Brillouin zone: 
$\int_{BZ} d{\bf k} \cdots=\int_{-\pi}^{\pi} dk_x dk_y dk_z \cdots$. The critical temperature 
in the absence of disorder is denoted by $T_c^{(0)}$ and it is 
defined by the condition $\mu(T_c^{(0)}) \equiv \mu_c =E_0$, where 
$E_0=-6t$ is the ground-state energy:  
\begin{equation}
\int_{BZ} \frac{d{\bf k}}{(2 \pi)^3}\frac{1}{e^{\beta_c^{(0)}
(E_{{\bf k}}-\mu_c)}-1}= f
\label{condizione-Tc}
\end{equation}
with $\beta_c^{(0)}=1/k_B T_c^{(0)}$.

A simple estimate of $T_c^{(0)}$ may be obtained by keeping 
only the lowest order of the Taylor expansion of the exponential in Eq.~(\ref{condizione-Tc}). One gets
\begin{equation}
\frac{1}{\beta_c^{(0)}} 
\int_{BZ} \frac{d{\bf k}}{(2 \pi)^3}\frac{1}{E_{{\bf k}}-\mu_c} \simeq f
\label{condizione-Tc-appr}
\end{equation}
yielding 
\begin{equation}
k_B T_c^{(0)} \simeq \frac{6tf}{{\sf W}(1)}.
\label{Tc0}
\end{equation}
In Eq.~(\ref{Tc0}) ${\sf W}(1) \simeq 1.516386$ and 
\begin{equation}
{\sf W}(z)\equiv 
\int_{BZ} \frac{d{\bf k}}{(2\pi)^3}\,\frac{1}{1-\frac{1}{3z}\sum_{\ell=1}^3\cos k_\ell} 
\label{watson}
\end{equation}
is the three-dimensional generalized Watson's integral \cite{mattis85}. 
Some useful properties of the generalized Watson's integral 
are grouped in Appendix A. 

In order to show that Eq.~(\ref{Tc0}) provides indeed an accurate estimate of 
$T_c^{(0)}$ for large filling, it is most convenient to write the exponential 
in Eq.~(\ref{condizione-Tc}) as a 
geometric series; this yields 
\begin{equation}
f=\sum_{j=1}^{\infty} e^{-6jX_0} \left[ I_0 \left( 2 j X_0 \right) \right]^3,
\label{alternative-condizione-Tc}
\end{equation}
where $X_0=\beta_c^{(0)}t$ 
and $I_0$ is the modified Bessel function 
of the first kind \cite{abramowitz64}. 
When $X_0 \ll 1$, one can approximate the series in (\ref{alternative-condizione-Tc}) with an integral and - for $f \gg 1$ - one obtains 
$f \simeq (1/6X_0) \int_0^\infty dx e^{-x}\left[I_0(x/3)\right]^3$. 
From Eq.~(\ref{int-rep-watson}), 
one readily gets Eq.~(\ref{Tc0}). 

Eq.~(\ref{alternative-condizione-Tc}) is very useful to numerically 
compute $X_0$ as a function of $f$ (or vice versa). 
Indeed, although the series in (\ref{alternative-condizione-Tc}) 
is slowly convergent, 
it is always possible to choose a large enough integer ${\cal N}$ so that 
Eq.~(\ref{alternative-condizione-Tc}) may be written as 
\begin{equation}
f=\sum_{j=1}^{{\cal N}} e^{-6jX_0} \left[ I_0 \left( 2 j X_0 \right) \right]^3+
{\cal R}({\cal N}),
\label{alternative-condizione-Tc_2}
\end{equation}
where the remainder of the series ${\cal R}$ may be approximated 
(with arbitrary precision for a pertinent choice of ${\cal N}$) as 
\begin{equation}
{\cal R}({\cal N}) \simeq \frac{1}{\left( 4 \pi X_0 \right)^{3/2}} \left( \zeta\left( \frac{3}{2} \right) -
\sum_{j=1}^{{\cal N}} \frac{1}{j^{3/2}}\right).
\label{remainder}
\end{equation}
In Eq.~(\ref{remainder}) the Riemann zeta function $\zeta$ 
and the asymptotic expansion of the modified Bessel function of the first kind 
$I_0(x) \simeq e^x/\sqrt{2 \pi x}$ valid for large $x$ \cite{abramowitz64} 
have been used. 

\begin{figure}[h]
\centering
\includegraphics[width=6cm,angle=270]{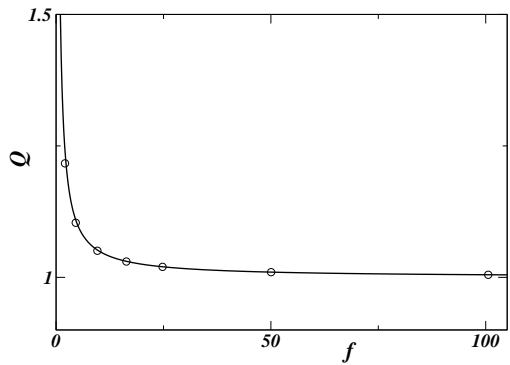}
\caption{Open circles: ratio $Q$ between the critical temperature $T_c^{(0)}$ 
and the large filling critical temperature
$\frac{6tf}{k_B {\sf W}(1)}$ as a function of the filling $f$.
Solid line:
expansion (\ref{correzioni_analitiche_3D}).}
\label{num-Tc0}
\end{figure}

In Fig.~\ref{num-Tc0} we plot the ratio $Q$ between the 
critical temperature $T_c^{(0)}$, determined numerically 
from Eqs.~(\ref{alternative-condizione-Tc_2}) and (\ref{remainder}), and 
the large filling critical temperature
$6tf / k_B {\sf W}(1)$ given by Eq.~(\ref{Tc0}), 
as a function of the filling $f$. The solid line
corresponds to the $1/f$ expansion, which is found to be
\begin{equation}
Q \equiv \frac{k_B T_c^{(0)}}{\frac{6tf}{{\sf W}(1)}} \approx 1+\frac{1}{2f}.
\label{correzioni_analitiche_3D}
\end{equation}
Notice that, at finite filling, $T_c^{(0)}$ is {\em larger} than
the critical temperature given for large filling in Eq.~(\ref{Tc0}).

We conclude this Section by observing that, using \cite{fetter71}
\begin{displaymath}
\lim_{\tau \to 0^-} \sum_{n=-\infty}^{\infty} \frac{e^{-i \omega_n \tau}}{i \omega_n-x}=
\frac{-\beta}{e^{\beta x}-1}
\end{displaymath}
with $\beta=\beta_c^{(0)}$ (the sum is over all the Matsubara frequencies $\omega_n=2 \pi n/ \beta_c^{(0)}$ 
and $\hbar=1$), Eq.~(\ref{condizione-Tc}) may be written as 
\begin{equation}
f=k_B T_c^{(0)}
\lim_{\tau \rightarrow 0^-} \sum_{n=-\infty}^{\infty} 
e^{-i\omega_n\tau} \int_{BZ}\frac{d{\bf k}}{(2\pi)^3}
\frac{1}{E_{{\bf k}}-\mu_c -i \omega_n},
\label{sum}
\end{equation}
which is valid for all values of $f$. From Eq.~(\ref{sum}), Eq.~(\ref{condizione-Tc-appr})
may be obtained by taking only the contribution from the lowest Matsubara frequency $\omega_0$. 
Contributions from higher Matsubara frequencies ($n=\pm1, \pm2, \cdots$) lead to a nonlinear equation relating  
$T_c^{(0)}$ and $f$. 

As we shall see, Eq.~(\ref{Tc0}) provides also the critical temperature of the SM discussed 
next.

\subsection{Correspondence with the Mean Spherical Model}

The Hamiltonian of the SM reads
\begin{equation}
H=-t\sum_{\langle i,j \rangle} S_i S_j,
\label{sph}
\end{equation}
where the values of the (classical) spins are real and 
satisfy the constraint $\sum_i S_i^2=N_S$. 
In the original solution of the SM \cite{berlin52}, the constraint 
is taken into account by writing the partition function as 
$Z=\int \prod_i dS_i e^{-\beta H} \delta(\sum_i S_i^2-N_S)$. At variance, in the so-called 
{\em mean} SM the constraint is satisfied only on average: upon introducing 
the grand-canonical Hamiltonian $K=H-\mu(\sum_i S_i^2-N_S)$, one only requires 
$\langle \sum_i S_i^2\rangle=N_S$, where $\langle \cdots \rangle$ denotes the thermal average with respect 
to $K$. 

In order to compare the results of the mean SM 
with the ones of an IBG with definite filling $f$, it is most convenient to write the constraint as: 
\begin{equation}
\sum_i S_i^2=f N_S=N_T.
\label{sph-constraint}
\end{equation}
The grand-canonical Hamiltonian is
\begin{equation}
K=H-\mu \left( \sum_i S_i^2-f N_S \right),
\label{sph-K}
\end{equation}
while the partition function is given by
\begin{equation}
Z=\int \prod_j dS_j e^{-\beta K}.
\label{sph-Z}
\end{equation}
For a cubic three-dimensional lattice, 
performing the gaussian integration in Eq.~(\ref{sph-Z}) yields
\begin{equation}
Z=e^{-\beta \mu f N_S} \prod_{{\bf k}}\sqrt{\frac{2 \pi}{\beta \left( 
E_{{\bf k}}-2\mu \right)}}.
\label{sph-Z-3D}
\end{equation}
Minimizing the free energy $F=-k_B T \ln{Z}$ with respect to $\mu$ leads to 
\begin{equation}
1=\frac{k_B T}{f N_S} \sum_{{\bf k}} \frac{1}{E_{{\bf k}}-2 \mu},
\label{sph-Z-N}
\end{equation}
from which, in the thermodynamical limit, one gets 
\begin{equation}
f=k_B T \int_{BZ}\frac{d{\bf k}}{(2 \pi)^3} \frac{1}{E_{{\bf k}}-2 \mu}.
\label{sph-Z-TD}
\end{equation}
The critical temperature $T_c^{(0)}$ is given by the condition $\mu(T_c^{(0)})=E_0/2$ yielding 
\begin{equation}
k_B T_c^{(0)} = \frac{6tf}{{\sf W}(1)}.
\label{Tc0-sph}
\end{equation}
Eq.~(\ref{Tc0-sph}) coincides with (\ref{Tc0}), which has been obtained for the lattice IBG 
for large filling. 

\section{Infinite-Range Model}

In this Section we investigate the $D \to \infty$ limit of the boson-hopping 
Hamiltonian (\ref{HAM}) without and with random-bond disorder. The main advantage 
of the infinite-range model lies both on its solvability and on the insights it provides 
for the behaviour of a realistic three-dimensional lattice model.

\subsection{Without disorder}

Let us start by considering the infinite-range boson hopping model ($A_{ij}=1$ if $i\neq j$). 
Its Hamiltonian is given by
\begin{equation}
\hat{H}^{(0)}=-\frac{t}{N_S} \sum_{i \neq j} \hat{b}^{\dag}_i \hat{b}_j.
\label{HAM_ord_LR}
\end{equation}
The single-particle energies are straightforwardly computed: one finds  
a non-degenerate ground-state energy $E_0=-t(1-1/N_S)$ and an excited state 
$E_1=t/N_S$ with degeneracy $N_S-1$. 
Conservation of the total number of particles yields
\begin{equation}
N_T=\frac{1}{e^{\beta (E_0-\mu)}-1}+
\frac{N_S-1}{e^{\beta(E_1-\mu)}-1}.
\label{cond-IR}
\end{equation}
As usual, the critical temperature $T_c^{(0)}$ is obtained from the 
condition $\mu(T_c^{(0)})=E_0$ which, in the thermodynamical limit, 
yields
\begin{equation}
k_B T_c^{(0)}=\frac{t}{\ln{\left( 1+\frac{1}{f}\right)}}.
\label{Tc_ord_LR}
\end{equation}
For $f \gg 1$, Eq.~(\ref{Tc_ord_LR}) leads to 
\begin{equation}
k_B T_c^{(0)} \simeq tf.
\label{Tc_ord_LR_large}
\end{equation}

\begin{figure}[h]
\centering
\includegraphics[width=6cm,angle=270]{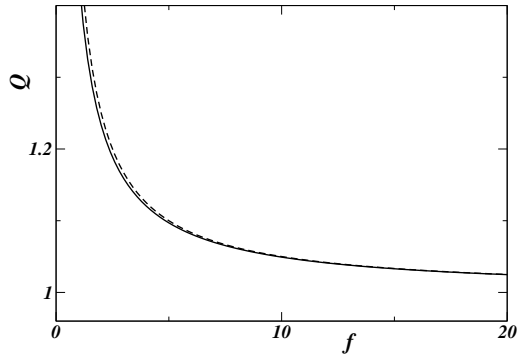}
\caption{Plot of the ratio $Q$ between the critical temperature 
$T_c^{(0)}$ and $tf/k_B$ as a function of the filling $f$ (solid line). The dashed line 
corresponds to the expansion at order $1/f$ given by 
Eq.~(\ref{correzioni_analitiche}).}
\label{LR}
\end{figure}

A plot of the ratio $Q$ between the critical temperature $T_c^{(0)}$ 
[given by Eq.~(\ref{Tc_ord_LR})] and the large filling 
critical temperature [given by Eq.~(\ref{Tc_ord_LR_large})] is provided 
in Fig.~\ref{LR}. The corrections to the large filling limit 
are of order $1/f$; indeed   
\begin{equation}
Q=\frac{k_B T_c^{(0)}}{tf}=\frac{1}{f \ln{\left( 1+\frac{1}{f}\right)}} \simeq 
1+\frac{1}{2f}.
\label{correzioni_analitiche}
\end{equation}
As for the three-dimensional lattice, 
the critical temperature at finite filling given by Eq.~(\ref{Tc_ord_LR}) is larger than 
the critical temperature determined for large $f$ in 
Eq.~(\ref{Tc_ord_LR_large}); it turns out that the $1/f$ correction has the same coefficient 
as in Eq.~(\ref{correzioni_analitiche_3D}). 
From Fig.~\ref{LR} one sees that the expansion (\ref{correzioni_analitiche}) 
is rather good also for intermediate filling, $f\gtrsim 1$.

To conclude this Section, we observe that 
if one adds a diagonal term to the Hamiltonian (\ref{HAM_ord_LR})
($A_{ii} = 1$), one still finds 
that the critical temperature is determined by Eq.~(\ref{Tc_ord_LR}). 

\subsection{With random-bond disorder}

In the presence of random-bond disorder the Hamiltonian is obtained from 
Eq.~(\ref{HAM_ord_LR}) replacing $A_{ij}$ with $\tilde{A}_{i,j}$. Namely,
\begin{equation}
\hat{H}=-\frac{t}{N_S} \sum_{i,j} (1+c_{ij})\, \hat{b}^{\dag}_i \hat{b}_j 
\equiv -\sum_{i,j} J_{ij} \hat{b}^{\dag}_i \hat{b}_j,
\label{HAM_ord_LR_dis}
\end{equation}
where, as in Ref.~\cite{kosterlitz76}, $J_{ij}$ is a real symmetric 
matrix with elements normally distributed, having average value $t/N_S$ and variance 
$J^2/N_S$ (but variance $2J^2/N_S$ on the diagonal). We set 
\begin{equation}
v=\frac{J}{t}
\label{def-v}
\end{equation}
and we limit ourself to $v<1$. The Hamiltonian (\ref{HAM_ord_LR_dis}) provides a mean-field theory description of the bond-disordered 
boson hopping Hamiltonian discussed in Section IV. 

In this Section we compute the dependence on the filling $f$ of the shift of the critical temperature, 
$\delta T_c \equiv T_c-T_c^{(0)}$, with $T_c^{(0)}$ defined in (\ref{Tc_ord_LR}). We find that while $\delta T_c<0$ 
for finite $f$, and, $\delta T_c \to 0$ for $f \to \infty$. 

\begin{figure}[h]
\centering
\includegraphics[width=6cm,angle=270]{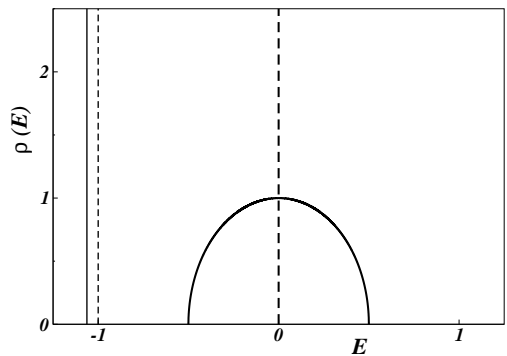}
\caption{Solid line: density of states $\rho(E)$ (in units of $1/\pi J$) as a function of $E$ 
(in units of $t$) for the infinite-range model with $v=0.25$ - 
the ground-state energy (\ref{e0-I-R}) is represented 
by a vertical solid line. The eigenvalues $E=0$ and $E=-t$ in the 
absence of disorder ($v=0$) are indicated by vertical dashed lines.}
\label{spectrum}
\end{figure}

In the infinite-range model the density of states can be obtained through an orthogonal
transformation mapping the problem in that of a single local impurity in a crystal
\cite{kosterlitz76,jones78}: one obtains the well-known semi-circular law \cite{mehta83}
and a single isolated impurity state which splits off from the continuous part of the spectrum. The same 
result can be obtained by the replica trick \cite{edwards76}. The density of states $\rho(E)$ is given by 
\begin{equation}
\rho(E)=\rho_0(E)+\frac{1}{N_S}\delta(E-E_0),
\label{rho-I-R}
\end{equation}
where 
\begin{equation}
E_0=-t(1+v^2).
\label{e0-I-R}
\end{equation}
The density of states $\rho_0(E)$ is given by $\rho_0(E)=0$ for $\mid E \mid > 2 J$ and 
\begin{equation}
\rho_0(E)=\frac{1}{2 \pi J^2} \sqrt{4J^2-E^2}
\label{rho0-I-R}
\end{equation}
for $\mid E \mid < 2 J$. A plot of the spectrum is provided in Fig.~\ref{spectrum}: one sees 
that the ground-state energy decreases 
(this induces an increase of the critical temperature $T_c$ for the bond-disordered model), while 
there is a spreading in the distribution of the continuous spectrum (instead, this 
induces a decrease of $T_c$). The shift in the critical temperature is a result of the competition 
of these two effects. 

Using Eq.~(\ref{rho-I-R}) one finds that the critical temperature is determined by 
($\beta_c=1/k_B T_c$): 
\begin{equation}
f = \frac{1}{2 \pi J^2} \int_{-2J}^{2J} dE \frac{\sqrt{4J^2-E^2}}{e^{\beta_c(E-E_0)}-1},
\label{ex-eq}
\end{equation}
with $E_0$ given by Eq.~(\ref{e0-I-R}). Defining $X=\beta_c t$ and 
$x = E/2J$, Eq.~(\ref{ex-eq}) becomes
\begin{equation}
f = \frac{2}{\pi} \int_{-1}^{1} dx \frac{\sqrt{1-x^2}}{e^{X(1+v^2)+2Xvx}-1}.
\label{ex-eq-X}
\end{equation} 
In order to compute $\delta T_c$ for $f \gg 1$, 
one can, as in Section II, Taylor expand the exponential in the r.h.s. of Eq.~(\ref{ex-eq-X}) to get 
\begin{equation}
f \simeq \frac{1}{\pi \beta_c t v} \int_{-1}^{1} dx \frac{\sqrt{1-x^2}}{x+(1+v^2)/2v} = \frac{1}{\beta_c t}.
\label{ex-eq-X-appr}
\end{equation}
From Eq.~(\ref{Tc_ord_LR_large}), one immediately sees that $\delta T_c \simeq 0$. 

For any finite value of the filling $f$ and for small $v$, $\delta T_c$ may be computed from the series expansion 
of the integrand of Eq.~(\ref{ex-eq-X}); one obtains 
\begin{equation}
f \simeq \frac{1}{e^X-1}+v^2 \frac{Xe^X}{(e^X-1)^2} \left( \frac{Xe^X}{e^X-1}-
\frac{X}{2}-1\right).
\label{ex-eq-X-dis}
\end{equation}
Upon defining $X_0=\beta_c^{(0)}t=\ln{(1+1/f)}$ and putting $X=X_0+\delta X$, one may readily linearize Eq.~(\ref{ex-eq-X-dis}) around 
$X_0$. One finds
\begin{equation}
\delta X \simeq v^2 X_0 \left( \frac{X_0 e^{X_0}}{e^{X_0}-1}-
\frac{X_0}{2}-1\right),
\label{ex-eq-X-dis-2}
\end{equation}
from which one gets
\begin{equation}
\frac{\delta T_c}{T_c^{(0)}} \simeq - v^2 \left[ \left( f+\frac{1}{2} \right) 
\ln{\left( 1+\frac{1}{f}\right)}-1\right].
\label{ex-eq-X-ris}
\end{equation}
Thus $\delta T_c<0$. 

\begin{figure}[h]
\centering
\includegraphics[width=6cm,angle=270]{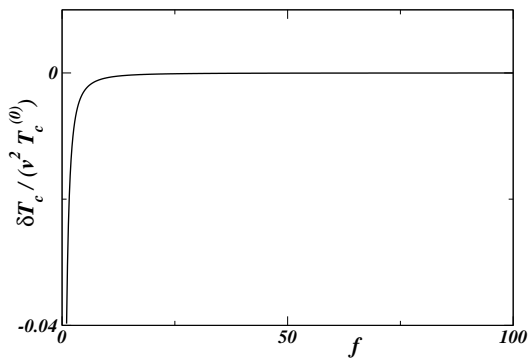}
\caption{Shift of the critical temperature (in units of $v^2 T_c^{(0)}$) 
as a function of the filling according Eq.~(\ref{ex-eq-X-ris}). 
$T_c^{(0)}$ is the critical temperature for the lattice without bond-disorder 
and it is given by Eq.~(\ref{Tc_ord_LR}).}
\label{v2-IR}
\end{figure}

\begin{figure}[h]
\centering
\includegraphics[width=6cm,angle=270]{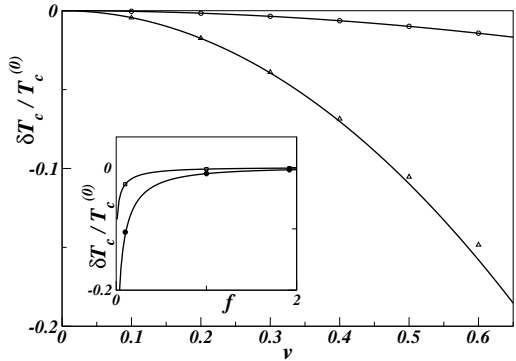}
\caption{Shift of the critical temperature $\delta T_c$ (in units of 
$T_c^{(0)}$) 
as a function of the disorder $v$: circles (triangles) are obtained 
from the numerical solution of Eq.~(\ref{ex-eq}) for $f=1$ 
($f=0.1$) - solid lines: analytical prediction Eq.~(\ref{ex-eq-X-ris}). 
Inset: $\delta T_c/T_c^{(0)}$ from Eq.~(\ref{ex-eq-X-ris}) vs. the filling $f$ for $v=0.25$ (squares) 
and $v=0.5$ (filled circles) - solid lines: Eq.~(\ref{ex-eq-X-ris}).}
\label{comparison-IR}
\end{figure}

Eq.~(\ref{ex-eq-X-ris}) is plotted in Fig.~\ref{v2-IR}. 
In Fig.~\ref{comparison-IR} we compare the result for $\delta T_c$ obtained 
from Eq.~(\ref{ex-eq-X-ris}) with the numerical solution 
of Eq.~(\ref{ex-eq}) for different values of the filling and the disorder. Although 
Eq.~(\ref{ex-eq-X-ris}) is valid for small $v$, a rather good agreement is found also for relatively large 
values of the disorder parameter $v$.  

In order to compute $f$ to any order in $v$, one observes that Eq.~(\ref{ex-eq-X}) can be written as
\begin{equation}
f=2\sum_{p=1}^\infty \frac{I_1(2vXp)}{2vXp} e^{-X(1+v^2)p},
\label{eq-I1}
\end{equation}
where $I_1$ is a modified Bessel function of the first kind 
\cite{abramowitz64}. Expanding both $I_1(2vXp)$ and $e^{-X(1+v^2)p}$ in terms of $v$ and using the definition of polylogarithm 
function, $Li_n(x)=\sum_{p=1}^\infty \frac{x^p}{p^n}$, one gets
\begin{equation}
f=\sum_{k,n=0}^{\infty}\left[\frac{(-1)^n X^{2k+n}}{n!\,k!\,\Gamma(k+2)}Li_{-(2k+n)}(e^{-X})\right]\,v^{2k+2n},
\label{ex-eq-X-dis-full}
\end{equation}
where $\Gamma$ is the Gamma function. 
It is easy to verify that, up to the order $v^2$, Eq.~(\ref{ex-eq-X-dis-full}) reduces to Eq.~(\ref{ex-eq-X-dis}).

We conclude this Section by noticing that, for the $D \to \infty$ SM, the effect of random-bond disorder 
has been worked out long ago \cite{kosterlitz76}: for disorder smaller than a critical value, i.e. 
for $v<1$, 
the authors of Ref.~\cite{kosterlitz76} found that the critical temperature $T_c$ of the
ferro-paramagnetic transition does not get shifted as a result of bond-disorder. Our result 
remarkably parallels the result obtained in Ref.~\cite{kosterlitz76}. 

\section{Three-Dimensional Hopping Bosons with Bond and On-Site Disorder}

In this Section we determine $\delta T_c$ for non-interacting bosons in $D=3$ with random-bond and on-site disorder. 
The Hamiltonian is given in Eq.~(\ref{HAM}), where $c_{ij}$ 
($\epsilon_i$) is a gaussian random variable 
with zero average and variance $v^2$ ($v_o^2t^2$). In the following we use the replica trick \cite{mezard87,dedominicis06}: 
we shall determine the Green functions of the disordered system from
the replicated action, where disorder enters as an
effective attractive interaction between replicas. As usual,
one introduces $N$ replicas and, after averaging the partition function on the
disorder, perform the limit $N \to 0$.
In the following, for the sake of simplicity, we
separately consider the effect of random-bond and on-site disorder.

\subsection{Random-Bond Disorder}

\begin{figure}[h]
\centering
\includegraphics[width=6cm]{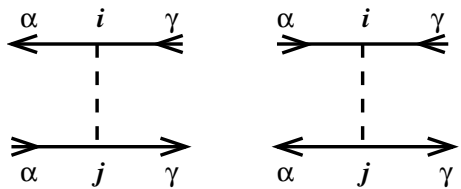}
\caption{Two-particle vertices for the disorder induced effective interaction.}
\label{fig1}
\end{figure}

A standard analysis (reported in Appendix B) allows us, after integrating over disorder, to derive an effective replica action as 
\begin{eqnarray}
\label{effective-action}
S_{eff}&=&
\sum_{\alpha} \int_{0}^{\beta} d \tau 
\left\{ \sum_i \varphi_i^{*\alpha}(\tau) 
\left( \frac{\partial}{\partial \tau} - \mu \right) \varphi_i^{\alpha}(\tau)  - 
t \sum_{ij} \varphi_i^{*\alpha}(\tau) A_{ij} \varphi_j^{\alpha}(\tau) \right\} \\
\nonumber&&-\frac{v^2t^2}{4}\sum_{ij\alpha\gamma}A_{ij}\int_0^\beta d\tau \int_0^\beta d\tau^\prime\Big(\varphi^{* \alpha}_i(\tau) \varphi^{\alpha}_j(\tau) \varphi^{* \gamma}_i(\tau^\prime) \varphi^{\gamma}_j(\tau^\prime)
+\varphi^{* \alpha}_i(\tau) \varphi^{\alpha}_j(\tau) 
\varphi^{* \gamma}_j(\tau^\prime) \varphi^{\gamma}_i(\tau^\prime)\Big).
\end{eqnarray}
The fields $\varphi$'s represent Bose particles and $\alpha, \gamma=1,...,N$ are the replica indices.
As one can see from (\ref{effective-action}), bond-disorder induces an effective interaction with the two-particle vertex functions 
depicted in Fig.~\ref{fig1}. Eq.~(\ref{effective-action}) is derived under the assumption that 
$\tilde A_{ij}=\tilde A_{ji}$ for any realization of disorder. 

Defining the Green's functions
\begin{eqnarray}
\label{G1}
&&{\cal G}_{ij}(z)=\langle\varphi_i\varphi_j^*\rangle_{v=0}=\int_{BZ}\frac{d{\bf k}}{(2\pi)^3}\frac{e^{i {\bf k} \cdot(i-j)}}{E_{\bf k}-z}\\
&&{G}_{ij}(z)=\langle\varphi_i\varphi_j^*\rangle_{v\neq 0}
\label{G2}
\end{eqnarray}
with $z=\mu+i\omega_n$ and performing a perturbative evaluation of 
Eq.~(\ref{G2}) up to the first order in $v^2$, one gets for $N\rightarrow 0$ \cite{footnote},
\begin{eqnarray}
G_{lm} = {\cal G}_{lm}+\frac{v^2t^2}{2}\sum_{ij} A_{ij}\left[
{\cal G}_{li}\,{\cal G}_{ij}\,{\cal G}_{jm}
+ {\cal G}_{lj}\,{\cal G}_{jm}\,{\cal G}_{ii}\right]+O(v^4).
\label{final}
\end{eqnarray}
Upon introducing the self-energy function ${\sf \Sigma}({\bf k},z)$, one may write the Dyson equation in Fourier space as follows
\begin{equation}
G^{-1}({\bf k},z)={\cal G}^{-1}({\bf k},z)+[{\sf \Sigma}({\bf k},z)-\delta\mu],
\end{equation}
where
\begin{equation}
{\cal G}^{-1}({\bf k},z)=E_{\bf k}-z
\end{equation}
and $\delta\mu$, the shift of the chemical potential, defined as
\begin{equation}
\delta \mu\equiv {\sf \Sigma}({\bf k}=0,z=\mu).
\end{equation}
Bose-Einstein condensation occurs when 
\begin{equation}
G({\bf k},z)^{-1}\Big{|}_{{\bf k}=0,z=\mu}=-6t-\mu+[{\sf \Sigma}(0,\mu)-\delta\mu]=0
\label{cond-BEC}
\end{equation}
which is solved for
\[\mu_c=-6t;\]
thus, for bond-disorder, the chemical potential at $T_c$ is given by $\mu_c+\delta\mu_c=-6t+{\sf \Sigma}(0,-6t)$.

Upon Fourier transforming Eq.~(\ref{final}), a straightforward but lengthy computation yields the Fourier component of ${\sf \Sigma}$ 
up to the order $v^2$. One gets
\begin{equation}
\label{self}
{\sf \Sigma}({\bf k},z)=-v^2t^2\left[3\int_{BZ} \frac{d{\bf k}^\prime}{(2\pi)^3}
\,\frac{1}{E_{{\bf k}^\prime}-z}
+\int_{BZ} \frac{d{\bf k}^\prime}{(2\pi)^3}\,\frac{\sum_{i=1}^3\cos(k^\prime_i-k_i)}{E_{{\bf k}^\prime}-z}\right].  
\end{equation}
Elementary computations based on the definition of 
Watson integral (see Eq.~(\ref{watson})) show that, introducing the functions
\begin{eqnarray}
\label{S1}
&&{\sf S}_1(z)\equiv 3t^2\int_{BZ} \frac{d{\bf k}^\prime}{(2\pi)^3}\,\frac{1}{E_{{\bf k}^\prime}-z}=-\frac{3t^2}{z}\,{\sf W}(-z/6t),\\
&&{\sf S}_2(z)\equiv-\frac{t}{2}\int_{BZ} 
\frac{d{\bf k}^\prime}{(2\pi)^3}\,\frac{E_{{\bf k}^\prime}}{E_{{\bf k}^\prime}-z}=-\frac{t}{2}\Big(1-{\sf W}(-z/6t)\Big)
\label{S2}
\end{eqnarray}
and using trigonometric relations, Eq.~(\ref{self}) may be compactly presented as 
\begin{equation}
\label{self_simple}
{\sf \Sigma}({\bf k},z)=-v^2\left({\sf S}_1(z)-\frac{1}{6t} {\sf S}_2(z)E_{\bf k}\right).
\end{equation}

From Eq.~(\ref{self_simple}) one may readily compute 
$\delta \mu_c$, as well as 
the renormalization of the bandwidth, which, {\it at the critical point}, are given by
\begin{eqnarray}
\delta\mu_c&=&-v^2\left({\sf S}_1(-6t)+{\sf S}_2(-6t)\right)\,=\,v^2(2{\sf W}(1)-1)t/2\approx v^2 t, \\
t_c&=&t+v^2\frac{1}{6}{\sf S}_2(-6t)\,=\,t\left[1+\frac{1}{12}v^2({\sf W}(1)-1)\right]\approx t(1+\frac{1}{24} v^2).
\label{bandwith}
\end{eqnarray}
Eq.~(\ref{bandwith}) shows that, as a result of the bond disorder, the bandwidth is increased: this is consistent with the mean-field 
result obtained in Section III.

The starting point of the computation of $\delta T_c$ is given by Eq.~(\ref{sum}). 
One notices that, as a result of the contribution in Eq.~(\ref{cond-BEC}) 
of the self-energy induced by the random-bond disorder, the filling fraction $f$ is changed to $f+\delta f$, with 
$\delta f$ given by
\begin{eqnarray}
\nonumber\delta f &=& \lim_{\tau\rightarrow 0^-} k_B T_c\sum_n e^{-i\omega_n\tau}
\int_{BZ} \frac{d{\bf k}}{(2\pi)^3} \left\{\frac{1}{E_{\bf k}-\mu_c-i\omega_n+{\sf \Sigma}({\bf k},\mu_c+i\omega_n)-\delta\mu_c}- \frac{1}{E_{\bf k}-\mu_c-i\omega_n}\right\}\\
&\simeq& -\lim_{\tau\rightarrow 0^-}k_B T_c\sum_n e^{-i\omega_n\tau}
\int_{BZ} \frac{d{\bf k}}{(2\pi)^3} \frac{\left[{\sf \Sigma}({\bf k},\mu_c+i\omega_n)-\delta\mu_c\right]}{(E_{\bf k}-\mu_c-i\omega_n)^2}.
\label{deltan}
\end{eqnarray}
Substituting 
\begin{eqnarray}
\label{I1}
\int_{BZ} \frac{d{\bf k}}{(2\pi)^3}\,\frac{1}{(E_{\bf k}-z)^2}&=&\frac{1}{3t}\,\frac{\partial}{\partial z} {\sf S}_1(z)\\
\label{I2}
\int_{BZ} \frac{d{\bf k}}{(2\pi)^3}\,\frac{E_{{\bf k}}}{(E_{\bf k}-z)^2}&=&-\frac{2}{t}\,\frac{\partial}{\partial z} {\sf S}_2(z)
\end{eqnarray}
in Eq.~(\ref{deltan}) one gets
\begin{equation}
\delta f = \lim_{\tau\rightarrow 0^-}k_B T_c\sum_n e^{-i\omega_n\tau}\frac{v^2}{3t^2}\left\{\Big[{\sf S}_1(z)-{\sf S}_1(\mu_c)\Big]\frac{\partial}{\partial z} {\sf S}_1(z)
+{\sf S}_2(z)\,\frac{\partial}{\partial z} {\sf S}_2(z)-{\sf S}_2(\mu_c)\,\frac{\partial}{\partial z} 
{\sf S}_1(z)\right\}\Big{|}_{z=\mu_c+i\omega_n}.
\label{dn_h}
\end{equation}

For large values of the filling $f$, the dominant contribution in Eq.~(\ref{dn_h})
is given by the lowest Matsubara frequency. 
The explicit computation carried in Appendix C yields
\begin{equation}
\delta f=-\left(\frac{9}{4\pi^2}+\frac{1}{9}{\sf W}(1)\Big({\sf W}(1)-1\Big)\right)\frac{v^2}{8t}k_B T_c\approx -0.04\frac{v^2}{t} k_B T_c.
\label{df}
\end{equation}
As a result of Eq.~(\ref{Tc0}), one notices that, for large $f$, there is a linear relation between $T_c^{(0)}$ and $f$, 
which is given by
\begin{equation}
\frac{{\sf W}(1)} {  6 t} \, k_B T_c^{(0)}=f.
\label{equating-1}
\end{equation}
At the order $v^2$, taking into account the 
shift in the particle density given by Eq.~(\ref{df}), one also has
\begin{equation}
\frac{{\sf W}(1) } {  6 t}   \, k_B T_c+\delta f=f.
\label{equating-2}
\end{equation}
From Eqs.~(\ref{equating-1}) and (\ref{equating-2}) and using Eq.~(\ref{df}), 
one obtains 
\begin{equation}
\label{3D_Tc(v)}
T_c=T_c^{(0)}\left\{
1+\frac{3}{4}v^2\left(\frac{9}{4\pi^2{\sf W}(1)}+\frac{1}{9}\Big({\sf W}(1)-1\Big)\right)\right\}\approx T_c^{(0)}\left(1+0.16\,v^2\right),
\end{equation}
implying that
\begin{equation}
\label{3D_dTc(v)}
\frac{\delta T_c}{T_c^{(0)}}\approx 0.16\,v^2. 
\end{equation}
${\delta T_c}/{T_c^{(0)}}$ is then positive and independent on the filling, provided that $f \gg 1$. 

We numerically checked the validity of Eq.~(\ref{3D_dTc(v)}). 
We adopted the following procedure. For several values of $v$ and $f$ and for several ($100$) 
configurations of hopping disorder and finite sized samples 
(cubic lattice with $L$ sites per side with $L=5,...,10$), 
we diagonalized exactly the Hamiltonian $H(L)$, 
then we averaged the single-particle eigenvalues over all the configurations, 
computing $T_c(L)$. 
Finally, we performed a finite-size scaling analysis using 
the equation $\frac{1}{T_c(L)}-\frac{1}{T_c}=\frac{c}{L}$ \cite{chamati96} in order to fit the parameters $c$ and $T_c\equiv T_c(\infty)$. 

\begin{figure}[h]
\centering
\includegraphics[width=10cm]{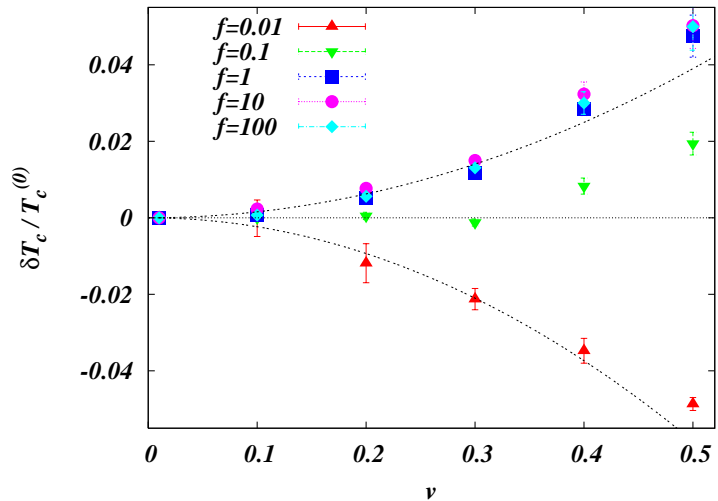}
\caption{Shift of the critical temperature $\delta T_c$ (in units of the 
critical temperature of the ordered lattice) as a function of the 
disorder strength $v$. The dots are obtained by numerical analysis 
while the two dash-lines are the analytic results: respectively, from the top 
of the figure, our result (\ref{3D_dTc(v)}) valid for large $f$ and the 
result of \cite{lopatin02} for the continuous disordered Bose gas.}
\label{Tc_v}
\end{figure}

In Fig.~\ref{Tc_v} the numerical results obtained for different values of $v$ and $f$
are compared with the analytic result (\ref{3D_dTc(v)}). 
One sees that, for small disorder and $f\gtrsim 1$, there is a good agreement between the two results. 
For $f$ very small, instead, we recover the result obtained by Lopatin and Vinokur \cite{lopatin02}: 
indeed, for small filling, the lattice details become irrelevant and one may treat the problem as in the continuum limit. 
Using the results of \cite{lopatin02} 
with $t\sim \hbar^2/2ma_0^2$ ($a_0$ being the lattice constant), one has 
$\frac{\delta T_c}{T_c^{(0)}}\approx -0.05\, \frac{v^2}{f^{1/3}}$. 
The resulting curve for $f=0.01$ is in very good agreement with the numerical results. 
The reason for this agreement is that, for a small filling, the occupied single-particle states lie at the bottom of the first energy band rendering 
the lattice ideal Bose gas basically equivalent to the continuous model.

In Appendix D we consider also the effect of an asymmetric random-bond 
disorder ($c_{ij}\neq c_{ji}$): the final result 
(\ref{Tc-asymm}) is similar to 
(\ref{3D_dTc(v)}), with a different 
numerical coefficient.

\subsection{Effects of on-site disorder}

It is easy to generalize our results to situations where also on-site disorder is present, see Eq.~(\ref{HAM}). 
Introducing the parameter $\gamma$, determined by the delta-correlated on-site random energies $\epsilon_i$ with variance $v^2_ot^2$, as 
\begin{equation}
\gamma\equiv\frac{3v^2+v^2_o}{3v^2},
\end{equation}
one gets that Eq.~(\ref{self_simple}) is modified as
\begin{equation}
\label{self_simple+}
{\sf \Sigma}({\bf k},z)=-v^2\left(\gamma\,{\sf S}_1(z)-\frac{1}{6t} {\sf S}_2(z)E_{\bf k}\right).
\end{equation}

When only on-site disorder is present ($v^2\rightarrow 0$), one has that 
${\sf \Sigma}({\bf k},z)=-(v^2_o/3) {\sf S}_1(z)$: this signifies that there is no renormalization 
of the bandwidth induced by a small random on-site disorder. 
Eq.~(\ref{self_simple+}) amount to state that the first term in Eq.~(\ref{dn_h}) should be multiplied by $\gamma$. 
As a result, when both random hopping and on-site disorder are present, the critical temperature 
is given by
\begin{equation}
T_c=T_c^{(0)}\left\{1+\frac{3}{4}v^2\left(\frac{6+3\gamma}{4\pi^2{\sf W}(1)}+\frac{1}{9}\Big({\sf W}(1)-1\Big)\right)\right\}\approx T_c^{(0)}\left(1+0.16\,v^2+0.01\,v^2_o\right).
\label{both}
\end{equation}
For comparable disorder strengths, $T_c$ is more sensitive to random-bond disorder, since the numerical coefficient 
of $v_o^2$ in Eq.~(\ref{both}) is smaller than the one associated to $v^2$. 
As a result, the dependence of $T_c$ on a small on-site disorder is extremely weak. We checked numerically 
that, for $v=0$, $v_o \lesssim 1$ and large $f$, $\delta T_c$ is very small. At variance, 
when $f \ll 1$, $\delta T_c$ is negative in agreement with the results of Ref.~\cite{lopatin02}. 

\subsection{Incommensurate potentials}

An experimental strategy employed to simulate the effect 
of disorder is based to the use a bichromatic incommensurate lattice 
\cite{fallani07,roati08}. The ultracold atomic gas 
is subjected to a periodic 
potential of the form $V\left( {\bf r} \right)=V_{main}\left( {\bf r} \right)+
V_{sec}\left( {\bf r} \right)$. While the main lattice potential creates 
the ordered lattice [e.g., an isotropic three-dimensional lattice corresponding to 
$V_{main}\left( {\bf r} \right)=V_0 
\left( \sin^2{\left( k_0 x \right)} + 
\sin^2{\left( k_0 y \right)}+ \sin^2{\left( k_0 z \right)}\right)$], the addition of 
a secondary lattice, with a spacing incommensurate with the one of the main 
lattice, leads to the emergence of a quasi-random potential which, in a finite sample, 
is equivalent, for all practical purposes, to a random potential. If the secondary potential 
is one-dimensional, $V_{sec}\left( {\bf r} \right)=V_1 
\sin^2{\left( k_1 x \right)}$, 
then, for large values of the height $V_0$ of the main lattice, 
one gets the Bose-Hubbard Hamiltonian (\ref{HAM}) 
with on-site energies $\epsilon_i$ of the form 
$\epsilon_i = - v \cos{\left( 2 \pi q i_x\right)}$, where 
$v$ depends on $V_1$ ($v \sim V_1/2$), $q=k_1/k_0$ and 
$i \equiv(i_x,i_y,i_z)$, with the $i_\ell$'s 
integers. In Ref.~\cite{fallani07}, the experimental values of 
$\lambda_0=2\pi/k_0$ and $\lambda_1=2\pi/k_1$ are 
$830nm$ and $1076 nm$, respectively.  
Of course, if the secondary lattice is three-dimensional,  
one would have $\epsilon_i \propto v \left( \cos{\left(2 \pi q i_x\right)}+
\cos{\left(2 \pi q i_y \right)} + 
\cos{\left(2 \pi q i_z \right)} \right)$. 

\begin{figure}[h]
\centering
\includegraphics[width=6cm,angle=270]{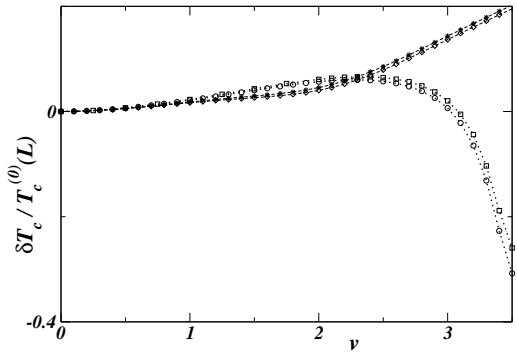}
\caption{Shift of the critical temperature (in units of the critical 
temperature $T_c^{(0)}(L)$ of the ordered lattice having a size $L$) 
as a function 
of $v$ for a bichromatic incommensurate lattice. Open circles (open squares): 
$f=1$ ($f=10$) for $q=0.206$. Diamonds (stars): $f=1$ ($f=10$) for 
$q=830/1076$. In both cases $L=10$. 
Dotted lines (for $q=0.206$) and dashed lines 
(for $q=830/1076$) are just a guide for the eye.}
\label{incomm}
\end{figure}

In Fig.~\ref{incomm} we plot, for two values of $q$, the shift in $T_c$  
for a finite size system ($L=10$) as a function of the disorder intensity 
$v$ of a one-dimensional secondary lattice 
for two different values of the filling: we see that, for small $v$, 
$\delta T_c>0$ while, for intermediate $v$, 
$\delta T_c$ may become negative. For small $v$, $\delta T_c \propto v^2$.

\section{Conclusions}

We investigated how the critical temperature for condensation of an ideal Bose gas on a cubic lattice is shifted as a consequence of random-bond 
and on-site disorder. Our analysis evidences how the shift in the critical temperature, $\delta T_c$, depends on the filling 
fraction $f$ and on the space dimension $D$. 

For the infinite-range model ($D \to \infty$), using results from
the theory of random matrices, we found the explicit dependence on $f$ of $\delta T_c$: $\delta T_c$ is negative 
and vanishes for large $f$, in agreement with known results
for the SM \cite{kosterlitz76}. 

For $D=3$, we wrote down the corrections to the Green functions due to disorder, finding that, 
for large $f$, $\delta T_c/T_c^{(0)}$ is positive and does not sensibly depend on $f$. 
At variance, for small $f$, $\delta T_c$ is negative and our results are in a very good agreement
with the ones obtained in Ref.~\cite{lopatin02} for the continuous disordered
Bose gas. The reason for this agreement is that, for small filling (i.e, much smaller
than one particle per site), the occupied single-particle states lie at the bottom of the first energy band: this renders 
the lattice ideal Bose gas basically equivalent to the continuous model investigated in Ref.~\cite{lopatin02}.

We presented numerical results 
for the shift of the critical temperature induced by the disorder generated by a bichromatic incommensurate 
potential in a realistic finite size system: we found that the critical 
temperature increases for small values of the height $v$ of the 
secondary lattice (i.e., small 
disorder), and decreases for larger values of $v$. The dependence 
of the shift of the critical temperature upon $v$ is $\delta T_c \sim v^2$.

We observe that, in the replicated action, the disorder enters as a fictitious 
effective attractive interaction between replicas. From this point of view, some
similarity emerges with the problem of determining the shift of
the critical temperature in a Bose gas due to the presence of a repulsive interaction 
\cite{baym99,holzmann01,kastening04,zobay06} (for a clear
presentation of this issue and more references, see \cite{blaizot08}): rather counterintuitively, the critical temperature increases
for small repulsive interactions,
while decreases for larger interaction. This should be compared with the results obtained 
for an homogeneous continuous Bose gas with weak disorder
\cite{lopatin02,zobay06,falco07}, where $T_c$ decreases with disorder, as a consequence of the 
inter-replicas attraction. For the ideal Bose gas we showed in this paper that, on a lattice, this result is modified: 
indeed, for small filling, 
only the bottom part of the energy spectrum is occupied and the effective attraction among replicas 
amounts to a decrease of the critical temperature, in agreement with \cite{lopatin02}. At variance, for large 
filling, all the single-particle energy band is occupied, including states with effective negative mass, leading to 
a net increase of $T_c$. 

We have considered non-interacting bosons on lattices 
with disorder: adding a repulsive interaction among the bosons, 
and treating it within the same approach we used, one expects \cite{blaizot08} that $T_c$ is unchanged. 
Going beyond the mean-field approach of this paper should unveil the interplay 
between the discreteness induced by the lattice and the interaction in a 
disordered environment. For this purpose, non-perturbative resummation techniques to compute 
the self-energy, such as the one discussed in Ref.~\cite{saavedra07}, could be relevant to determine $\delta T_c$. 
In the same way, it would be very interesting to study the 
effects of an harmonic confinement on the disordered lattice \cite{timmer06}. 

To conclude, we mention that if one introduces suitable time-dependent 
optical lattices \cite{eckardt05,sias08}, 
one should be able to see the effects also of negative effective $t_{ij}$'s: for non-interacting bosons, 
this could provide an experimental testing ground for the three-dimensional spin glass spherical model.

{\em Acknowledgements:} Discussions with 
D. M. Basko, V. E. Kravtsov, S. Giorgini, 
F. Becca, M. Fabrizio, G. Modugno, and G. Mussardo are warmly acknowledged. 
P.S. thanks SISSA for hospitality at several stages of this work. 
L.D.A., P.S. and A.T. thank the Galileo Galilei Institute for Theoretical 
Physics for the hospitality and INFN for partial support during the completion 
of this work. This work is partly supported by the MIUR project 
``Quantum Noise in Mesoscopic Systems''.

\appendix

\section{Some useful properties of the Watson integrals}

The generalized Watson's integrals are defined by 
\begin{equation}
{\sf W}(d,z)=\frac{1}{\left( 2 \pi \right)^d} \int_{-\pi}^{\pi} \prod_{i=1}^{d}
dk_i \frac{1}{1-\frac{1}{dz} \sum_{j=1}^{d} \cos{k_j}};
\label{watson-gen}
\end{equation}
in our paper we are concerned only with ${\sf W}(z) \equiv {\sf W}(3,z)$. 

The Watson integral (\ref{watson}) admits an integral representation 
\begin{equation}
{\sf W}(z)=\int_0^\infty dx \,e^{-x}\left[I_0(x/3z)\right]^3
\label{int-rep-watson}
\end{equation}
in terms of the modified Bessel function of the first kind $I_0$. 
Particularly important is its value at $z=1$, which is given by
${\sf W}(1)= \frac{\sqrt{6}}{32 \pi^3}
\Gamma\left(\frac{1}{24}\right) \Gamma\left(\frac{5}{24}\right)
\Gamma\left(\frac{7}{24}\right) 
\Gamma\left(\frac{11}{24}\right)\simeq 1.516386$ \cite{joyce72-b}.

The Watson function ${\sf W}(z)$ can also be written as a probability generating function
\begin{equation}
\label{sumP}
{\sf W}(z)=\sum_{n=0}^{\infty} \frac{P_n}{z^n}, \;\;\; |z|\ge 1 
\end{equation}
with $P_n$ being the probability that a random walker returns after $n$ steps at the same point of the cubic lattice. 
In particular, 
\begin{equation}
P_{2n}=\int_{BZ} \frac{d{\bf k}}{(2\pi)^3} \left(\frac{1}{3}\sum_{\ell=1}^3\cos k_\ell\right)^{2n}
\end{equation}
is the probability for a random walker on a cubic lattice of reaching 
the starting site after $2n$ steps. 
Clearly, $P_{2n+1}=0$ since a closed loop on a square lattice has always an even number of segments. 
For this reason one can write ${\sf W}(-z/6t)$, which is needed in the computation of Section 
IV and Appendix C, as 
\begin{equation}
\label{sumP2}
{\sf W}(-z/6t)=\sum_{n=0}^\infty \frac{(-6t)^{2n} P_{2n}}{z^{2n}}.
\end{equation}
We notice that ${\sf W}(z)$ is also the solution of a Riemann p-differential equation, so that, when
$|z|\ge 6t$, ${\sf W}(-z/6t)$ and its derivative are well defined.

\section{Effective replicated action}

Let us consider the Hamiltonian 
\begin{equation}
\hat{H}=-t\sum_{i,j} \tilde{A}_{ij} \hat{b}^{\dag}_i \hat{b}_j - 
\mu \sum_{i} \hat{b}^{\dag}_i \hat{b}_i  + 
\sum_i \epsilon_i \hat{b}^{\dag}_i \hat{b}_i
\label{HAM-app}
\end{equation}
where $\tilde{A}_{ij}=A_{ij} (1+c_{ij})$, with $A_{ij}=1,0$ depending on 
if $i$ and $j$ are neighbour sites or not, $c_{ij}$, $\epsilon_i$ being random variables. 

The partition function can be written as a coherent-state path integral  
(in units where $\hbar=k_B=1$)
\begin{equation}
Z=\int \prod_i {\cal D}\varphi_i {\cal D}\varphi_i^\ast e^{-S},
\label{Z-app}
\end{equation}
where $S$ is given by
\begin{equation}
S=\int_0^{\beta} d \tau \left\{ \sum_i \varphi_i^\ast 
\left( \frac{\partial}{\partial \tau} -\mu \right) \varphi_i +
\sum_i \epsilon_i \varphi_i^\ast \varphi_i -
t \sum_{i,j} \tilde{A}_{ij} \varphi^{\ast}_i \varphi_j
\right\}.
\label{action-app}
\end{equation}
The partition function (\ref{Z-app}) depends on the disorder realization of the hopping 
$\{c\}$ and of the on-site energies $\{\epsilon\}$: $Z=Z(\{c\},\{\epsilon\})$. Labeling the $N$ replicas by $\alpha=1,\cdots,N$ and using $\tilde{A}_{ij}=\tilde{A}_{ji}$, one has 
\begin{equation}
Z^N(\{c\},\{\epsilon\})=\int \prod_{i,\alpha} {\cal D}\varphi_i^\alpha {\cal D}\varphi_i^{\alpha \ast} 
e^{-\int_0^{\beta} d \tau \left\{ \sum_{i,\alpha} \varphi_i^{\alpha \ast} 
\left( \frac{\partial}{\partial \tau} -\mu +\epsilon_i \right) \varphi_i^\alpha -
\frac{t}{2} \sum_{i,j,\alpha} \tilde{A}_{ij} \left(\varphi^{\alpha \ast}_i \varphi_j^\alpha +\varphi^{\alpha \ast}_j \varphi_i^\alpha \right)
\right\}}.
\label{Z-app-rep}
\end{equation}

If, for simplicity, one assumes that $\epsilon_i=0$ (only random-bond disorder is present), the averaged effective partition function is
\begin{equation}
\overline{Z^N(\{c\})}=\int \prod_{i,j} dc_{ij} P(c_{ij}) Z^N(\{c\}),
\label{Z-app-rep-eps}
\end{equation}
where $P(c_{ij})$ is the gaussian probability distribution 
\begin{equation}
P(c_{ij})=\frac{1}{\sqrt{2\pi v^2}}e^{-(c_{ij})^2/2v^2},
\label{prob-gauss}
\end{equation}
with variance $v^2$. After integrating over the $c_{ij}$'s, one gets 
\begin{equation}
\overline{Z^N(\{c\})}=\int \prod_{i,\alpha} {\cal D}\varphi_i^\alpha {\cal D}\varphi_i^{\alpha \ast} 
e^{-S_{eff}},
\label{averaged}
\end{equation}
where $S_{eff}$ is given in Eq.~(\ref{effective-action}). 
When on-site disorder $\{\epsilon\}$ is present, one can follow the same procedure: namely, one averages Eq.~(\ref{Z-app-rep}) 
over disorder with the measure $\prod_{i}d\epsilon_i \,P_o(\epsilon_i)\equiv\prod_{i}d\epsilon_i \,\frac{1}{\sqrt{2\pi v_o^2 t^2}}
e^{-(\epsilon_i)^2/2v_o^2t^2}$, where $v_o^2t^2$ is the variance of on-site random energies ($v_o$ is a dimensionless quantity).

\section{Shift of the filling fraction}

Using Eqs.~(\ref{S1}),~(\ref{S2}),~(\ref{self_simple}),~(\ref{I1}) and (\ref{I2}) 
in order to write $\delta f$ only in terms of the Watson function ${\sf W}(-z/6t)$, one obtains Eq.~(\ref{dn_h}). 
In Eq.~(\ref{dn_h}) the dominant contribution to $\delta f$ is given by the lowest Matsubara frequency ($\omega_n=0$): 
\begin{equation}
\delta f=k_B T_c \frac{v^2}{3t^2}
\lim_{z\rightarrow \mu_c}
\left\{\Big[{\sf S}_1(z)-{\sf S}_1(\mu_c)\Big]\frac{\partial}{\partial z} {\sf S}_1(z)
+{\sf S}_2(z)\,\frac{\partial}{\partial z} {\sf S}_2(z)-{\sf S}_2(\mu_c)\,\frac{\partial}{\partial z} {\sf S}_1(z)\right\}.
\label{dn_h_0}
\end{equation} 
One can compute Eq.~(\ref{dn_h_0}) taking advantage of the asymptotic behavior of ${\sf W}(z)$. 
For $\eta^2\equiv z^2-(6t)^2\rightarrow 0$ one can write
\begin{equation}
{\sf W}(-z/6t)
={\sf W}(1)-\frac{3\sqrt{3}}{2\pi}\sqrt{1-\frac{(6t)^2}{z^2}}+
{\cal K}
\left(1-\frac{(6t)^2}{z^2}\right)
+O(\eta^3)
\end{equation}
where ${\cal K}=\frac{9}{32}\left({\sf W}(1)+\frac{6}{\pi^2{\sf W}(1)}\right)$. From Eqs.~(\ref{S1})-(\ref{S2}) one has 
\begin{eqnarray}
&&{\sf S}_1(z)= -\frac{3t^2}{z}\left({\sf W}(1)-\frac{3\sqrt{3}}{2\pi}\sqrt{1-\frac{(6t)^2}{z^2}}\right)+O(\eta^2),\\
\label{S1-ex}
&&{\sf S}_2(z)=-\frac{t}{2}\left(1-{\sf W}(1)+\frac{3\sqrt{3}}{2\pi}\sqrt{1-\frac{(6t)^2}{z^2}}\right)+O(\eta^2).
\label{S2-ex}
\end{eqnarray}
The derivative of these functions are
\begin{eqnarray}
\label{dS1}
&&\frac{\partial}{\partial z}{\sf S}_1(z)=3 t^2\left(\frac{3\sqrt{3}}{2\pi}\frac{(6t)^2}{z^4}\frac{1}{\sqrt{1-\frac{(6t)^2}{z^2}}}+\frac{W(1)}{z^2}-\frac{2(6t)^2{\cal K}}{z^4}\right)+O(\eta),\\
\label{dS2}
&&\frac{\partial}{\partial z}{\sf S}_2(z)=-\frac{t}{2}\left(\frac{3\sqrt{3}}{2\pi}\frac{(6t)^2}{z^3}\frac{1}{\sqrt{1-\frac{(6t)^2}{z^2}}}-\frac{2(6t)^3{\cal K}}{z^3}\right)+O(\eta).
\end{eqnarray}
One should notice that Eqs.~(\ref{dS1}, \ref{dS2}) diverge at $z=-6t$, but these divergences are canceled in Eq.~(\ref{dn_h_0}). 
Inserting Eqs.~(\ref{S1-ex})-(\ref{dS2}) in Eq.~(\ref{dn_h_0}), one readily finds 
Eq.~(\ref{df}).

An alternative way to obtain Eq.~(\ref{df}) is based 
on the random-walk probability expansion of the Watson integral presented 
in Appendix A. If one truncates the sum (\ref{sumP2}), 
the result for $\delta T_c$ would be zero since, only for the infinite sum, one has that  
\begin{equation}
\frac{\partial}{\partial z}{\sf W}(-z/6t)=-\sum_{n=1}^{\infty}\frac{n(-6t)^{2n} P_{2n}}{z^{2n+1}}
\label{der-app}
\end{equation}
is not convergent at $z=-6t$. 
Eq.~(\ref{der-app}) implies that only {\em very long random walks are important}: thus, 
expansion in terms of $n^{-1}$ 
(the inverse length of the walks) is meaningless since only walks with $n\rightarrow \infty$ are relevant, 
although $P_n$ decreases very fast with $n$. For large $n$, $P_{2n}$ has the following asymptotic behavior
\begin{equation}
\label{P_2n_asim}
P_{2n}\sim 2\left(\frac{3}{4\pi n}\right)^{3/2},\;\;n\rightarrow \infty.
\end{equation} 
Inserting Eq.~(\ref{P_2n_asim}) in Eq.~(\ref{sumP2}) (for $n\ge1$, and $P_0=1$) and using Eqs (\ref{S1}, (\ref{S2}) and (\ref{dn_h_0})), 
after taking the limit $z\rightarrow -6t$, one gets Eq.~(\ref{df}). 

\section{Asymmetric random-bond disorder}

Here we consider the situation where $c_{ij}$ and $c_{ji}$ are two
independent random variables, namely 
\begin{equation}
\label{ineq}
c_{ij}\neq c_{ji}.
\end{equation}
This situation may arise, for instance, when time reversal symmetry is broken. 

After integrating over disorder, one gets
\begin{equation}
G_{lm} = {\cal G}_{lm}+v_A^2t^2\sum_{ij} A_{ij}
(N+1){\cal G}_{li}\,{\cal G}_{ij}\,{\cal G}_{jm}+O(v_A^4)
\end{equation}
(where $v_A^2$ is the variance of the asymmetric random-bond disorder), 
which implies that 
\begin{equation}
{\sf \Sigma}({\bf k},z)=\frac{v_A^2}{3t} {\sf S}_2(z)E_{\bf k}.
\end{equation}
As a result, an asymmetric random-bond disorder yields a value of $T_c$ given by 
\begin{equation}
T_c=T_c^{(0)}\left\{1+\frac{3}{2}v_A^2\left(\frac{6}{4\pi^2{\sf W}(1)}+\frac{1}{9}\Big({\sf W}(1)-1\Big)\right)\right\}\approx 
T_c^{(0)}\left(1+0.24\,v_A^2\right).
\label{Tc-asymm}
\end{equation}

In order to convince oneself that time reversal symmetry could play a role, 
one may consider a system
with only two sites and random hopping between them, described by a matrix
\begin{eqnarray}
\left(
\begin{array}{cc}
0&t_{12}\\
t_{21}&0
\end{array}\right).
\label{eig-app}
\end{eqnarray} 
The eigenvalues of (\ref{eig-app}) are
\begin{equation}
\lambda_{\pm}=\pm \sqrt{t_{12}t_{21}}.\end{equation}
If time reversal symmetry holds, one has that 
\begin{equation} t_{12}=t_{21}. \end{equation} 
Averaging on the disorder and assuming $\overline{t_{12}}=\overline{t_{21}}=t$, one gets 
\begin{equation} \overline{\lambda_{\pm}}=\pm \,t.
\end{equation}
On the other hand, if time reversal symmetry is broken (i.e., $t_{12}\neq t_{21}$), 
one has 
\begin{equation} 
\overline{\lambda_{\pm}}= \pm \,t 
\left(\frac{(1+\varepsilon)^{3/2}-(1-\varepsilon)^{3/2}}{3\varepsilon}\right)^2\simeq \pm \,t 
\left(1-\frac{\varepsilon^2}{12}\right).
\label{average-dis}
\end{equation}
In deriving Eq.~(\ref{average-dis}) a uniform distribution for $t_{12}$ and $t_{21}$ with width $2t\varepsilon$ and mean value $t$ 
has been assumed.


\begin{thebibliography}{99}

\bibitem{landau80} L. D. Landau, E. M. Lifshits, and L. P. Pitaevskii, 
{\it Statistical physics}, Pergamon (London, 1980)

\bibitem{huang87} K. Huang, {\it Statistical mechanics}, Wiley (New York, 1987)

\bibitem{pathria96} R. K. Pathria, {\it Statistical mechanics}, Butterworth-Heinemann (Oxford, 1996)

\bibitem{pathria98} See a pedagogical treatment of the ideal Bose gas in a finite system in R. K. Pathria, Am. J. Phys. 
{\bf 66}, 1080 (1998) 

\bibitem{pethick02} C. J. Pethick and H. Smith, {\it Bose-Einstein condensation in dilute alkali gases},
Cambridge University Press (Cambridge, 2002)
 
\bibitem{stringari03} L. P. Pitaveskii and S. Stringari, {\it Bose-Einstein condensation}, Clarendon Press (Oxford, 2003)

\bibitem{jaksch98} D. Jaksch, C. Bruder, J. I. Cirac, 
C. W. Gardiner, and P. Zoller, Phys. Rev. Lett. {\bf 81}, 3108 (1998)

\bibitem{trombettoni01} A. Trombettoni and A. Smerzi, Phys. Rev. Lett.
{\bf 86}, 2353 (2001)

\bibitem{roati07} G. Roati, M. Zaccanti, C. D'Errico, J. Catani, M. Modugno, A. Simoni, M. Inguscio, 
and G. Modugno, Phys. Rev. Lett. {\bf 99}, 010403 (2007)   

\bibitem{nota-dip} A small amount of dipolar interaction is yet present, but one can further tune $a$ in order to partially
compensate such residual interaction, as discussed in 
M. Fattori {\it et al.}, arXiv:0808.1506 

\bibitem{gustavsson08} M. Gustavsson, E. Haller, M. J. Mark, J. G. Danzl, G. Rojas-Kopeinig, and 
H.-C. N\"agerl, Phys. Rev. Lett. {\bf 100}, 080404 (2008)
 
\bibitem{fattori08-b} M. Fattori, C. D'Errico, G. Roati, M. Zaccanti, M. Jona-Lasinio, 
M. Modugno, M. Inguscio, and G. Modugno, Phys. Rev. Lett. {\bf 100}, 080405 (2008)

\bibitem{lacour74} P. Lacour-Gayet and G. Toulouse, J. Phys. (Paris) {\bf 35}, 425 (1974)

\bibitem{roati08} G. Roati, C. D'Errico, L. Fallani, M. Fattori, C. Fort, M. Zaccanti, G. Modugno, M. Modugno, and M. Inguscio, Nature {\bf 453}, 895 (2008)  

\bibitem{billy08} J. Billy, V. Josse, Z. C. Zuo, A. Bernard, B. Hambrecht, P. Lugan, D. Cl\'ement, L. Sanchez-Palencia, P. Bouyer, and A. Aspect, 
Nature {\bf 453}, 891 (2008)  

\bibitem{lye05} J. E. Lye, L. Fallani, M. Modugno, D. S. Wiersma, C. Fort, and M. Inguscio, 
Phys. Rev. Lett. {\bf 95}, 070401 (2005) 

\bibitem{clement05} D. Cl\'ement, A. F. Var\'on, M. Hugbart, 
J. A. Retter, P. Bouyer, L. Sanchez-Palencia, D. M. Gangardt, G. V. Shlyapnikov, and A. Aspect, 
Phys. Rev. Lett. {\bf 95}, 170409 (2005) 

\bibitem{fallani07} L. Fallani, J. E. Lye, V. Guarrera, C. Fort, and M. Inguscio, Phys. Rev. Lett. {\bf 98}, 130404 (2007)  

\bibitem{giamarchi88} T. Giamarchi and H. J. Schulz, Phys. Rev. B  {\bf 37}, 325 (1988)

\bibitem{fisher89} M. P. Fisher, P. B. Weichman, G. Grinstein, 
and D. S. Fisher, Phys. Rev. B {\bf 40}, 546 (1989)

\bibitem{gimperlein05} H. Gimperlein, S. Wessel, J. Schmiedmayer, and 
L. Santos, Phys. Rev. Lett. {\bf 95}, 170401 (2005)  

\bibitem{rey06} A. M. Rey, I. I. Satija, and C. W. Clark, 
Phys. Rev. A {\bf 73}, 063610 (20060
 
\bibitem{krutitsky06} 
K. V. Krutitsky, A. Pelster, and R. Graham, New J. Phys. {\bf 8}, 187 (2006); 
V. I. Yukalov, E. P. Yukalova, K. V. Krutitsky, 
and R. Graham, Phys. Rev. A {\bf 76}, 053623 (2007); 
K. V. Krutitsky, M. Thorwart, R. Egger, and R. Graham, Phys. Rev. A {\bf 77}, 053609 (2008)  

\bibitem{roscilde07} T. Roscilde and J. I. Cirac, Phys. Rev. Lett. {\bf 98}, 190402 (2007); T. Roscilde, Phys. Rev. A {\bf 77}, 063605 (2008)
 
\bibitem{sengupta07} P. Sengupta and S. Haas, Phys. Rev. Lett. {\bf 99}, 050403 (2007) 
 
\bibitem{buonsante07} P. Buonsante, V. Penna, A. Vezzani, and P. B. Blakie, 
Phys. Rev. A {\bf 76}, 011602 (2007)

\bibitem{deng08} X. Deng, R. Citro, A. Minguzzi, and E. Orignac, 
Phys. Rev. A {\bf 78}, 013625 (2008) 

\bibitem{roux08} G. Roux, T. Barthel, I. P. McCulloch, C. Kollath, 
U. Schollw\"ock, and T. Giamarchi, Phys. Rev. A {\bf 78}, 023628 (2008)

\bibitem{morrison08} S. Morrison, A. Kantian, A. J. Daley, 
H. G. Katzgraber, M. Lewenstein, H. P. Buchler, and P. Zoller, 
New J. Phys. {\bf 10}, 073032 (2008)  

\bibitem{bissbort08} U. Bissbort and W. Hofstetter, arXiv:0804.0007

\bibitem{kiewiet75} C. W. Kiewiet, H. E. Hall, and J. D. Reppy, Phys. Rev. Lett. {\bf 35}, 1286 (1975)

\bibitem{hertz79} J. A. Hertz, L. Fleishman, and P. W. Anderson, 
Phys. Rev. Lett. {\bf 43}, 942 (1979)


\bibitem{huang92} K. Huang and H.-F. Meng, Phys. Rev. Lett. {\bf 69}, 644 (1992)

\bibitem{giorgini94} S. Giorgini, L. P. Pitaevskii, and S. Stringari, 
Phys. Rev. B {\bf 49}, 12938 (1994)

\bibitem{lopatin02} A. V. Lopatin and V. M. Vinokur, Phys. Rev. Lett. {\bf 88}, 235503 (2002)

\bibitem{yukalov07} V.I. Yukalov and R. Graham, Phys. Rev. A 
{\bf 75}, 023619 (2007)


\bibitem{gunton68} J. D. Gunton and M. J. Buckingham, Phys. Rev. {\bf 166}, 152 (1968)

\bibitem{berlin52} T. H. Berlin and M. Kac, Phys. Rev. {\bf 86}, 821 (1952)

\bibitem{joyce72} G. S. Joyce, in {\em Phase transitions and critical phenomena}, 
vol. 2, C. Domb and M. S. Green eds., p. 375, Academic Press (London, 1972)

\bibitem{mattis85} See Sections 9-13 in D. C. Mattis {\it The theory of magnetism II}, 
Springer-Verlag (Berlin, 1985)

\bibitem{joyce66} G. S. Joyce, Phys. Rev. {\bf 146}, 349 (1966)  

\bibitem{fisher72} M. E. Fisher and M. N. Barber, Phys. Rev. Lett. {\bf 28}, 1516 (1972)

\bibitem{stanley68} H. E. Stanley, Phys. Rev. {\bf 176} , 718 (1968)

\bibitem{brezin93} {\it The large N expansion in quantum field theory and statistical physics: 
from spin systems to 2-dimensional gravity}, E. Brezin and S. R. Wadia eds., World Scientific (Singapore, 
1993)

\bibitem{kosterlitz76}  J. M. Kosterlitz, D. J. Thouless, and 
R. C. Jones, Phys. Rev. Lett. {\bf 36}, 1217 (1976)

\bibitem{dedominicis06} C. De Dominicis and I. Giardina, {\it Random fields and spin glasses: 
a field theory approach}, Cambridge University Press (Cambridge, 2006)

\bibitem{pirc99} R. Pirc and R. Blinc, Phys. Rev. B {\bf 60}, 13470 (1999)  

\bibitem{abramowitz64} {\it Handbook of mathematical functions with formulas, graphs, and mathematical tables}, 
M. Abramowitz and I. A. Stegun eds.,  National Bureau of Standards (Washington, 1964)

\bibitem{fetter71} A. L. Fetter and J. D. Walecka, {\it Quantum theory of many-particle systems}, 
McGraw-Hill (New York, 1971)

\bibitem{jones78} R. C. Jones, J. M. Kosterlitz, and D. J. Thouless, 
J. Phys. A {\bf 11}, L45 (1978)  

\bibitem{mehta83} M. L. Mehta, {\it Random matrices}, Elsevier (Amsterdam, 1983).

\bibitem{edwards76} S. F. Edwards and R. C. Jones, J. Phys. A {\bf 9}, 1595 (1976)  

\bibitem{mezard87} M. Mezard, G. Parisi, and M. Virasoro, 
{\it Spin glass theory and beyond}, World Scientific (Singapore, 1987)


\bibitem{footnote} Notice that, before taking the limit, the expression for $G$ is analogous to the standard spinful case where instead of the spin $s$ we have $N$ flavors: 
$
\nonumber G_{lm} = {\cal G}_{lm}+\frac{v^2t^2}{2}\sum_{ij} A_{ij}\left[
(2N+1){\cal G}_{li}\,{\cal G}_{ij}\,{\cal G}_{jm}
+ {\cal G}_{lj}\,{\cal G}_{jm}\,{\cal G}_{ii}\right]+O(v^4)
$.

\bibitem{chamati96} H. Chamati and N. S. Tonchev, J. Stat. Phys. {\bf 83}, 1211 (1996)  


\bibitem{baym99} G. Baym, J.-P. Blaizot, M. Holzmann, F. Lalo\"e, and D. Vautherin, 
Phys. Rev. Lett. {\bf 83}, 1703 (1999)

\bibitem{holzmann01} M. Holzmann, G. Baym, J.-P. Blaizot, and F. Lalo\"e, 
Phys. Rev. Lett. {\bf 87}, 120403 (2001)

\bibitem{kastening04} B. Kastening, Phys. Rev. A {\bf 69}, 043613 (2004) 

\bibitem{zobay06} O. Zobay, Phys. Rev. A {\bf 73}, 023616 (2006)
 
\bibitem{blaizot08} J.-P. Blaizot, arXiv:0801.0009

\bibitem{falco07} G. M. Falco, A. Pelster, and R. Graham, Phys. Rev. A {\bf 75}, 063619 (2007) 

\bibitem{saavedra07} F. A. de Saavedra, C. Bisconti, G. Co', and A. Fabrocini, Phys. Rep. {\bf 450}, 1 (2007)   
 
\bibitem{timmer06} M. Timmer, A. Pelster, and R. Graham, 
Europhys. Lett. {\bf 76}, 760 (2006) 

\bibitem{eckardt05} A. Eckardt, J. T. Jinasundera, C. Weiss, and M. Holthaus, 
Phys. Rev. Lett. {\bf 95}, 200401(2005)  

\bibitem{sias08} C. Sias, H. Lignier, Y. P. Singh, 
A. Zenesini, D. Ciampini, O. Morsch, and E. Arimondo, 
Phys. Rev. Lett. {\bf 100}, 040404 (2008)  

\bibitem{joyce72-b} G. S. Joyce, J. Phys. A {\bf 5}, L65 (1972)


\end{thebibliography}
\end{document}